\documentclass[8.5pt,twoside,twocolumn]{article}
\usepackage[super,sort&compress,comma]{natbib}
\usepackage{mhchem}
\usepackage{times,mathptmx}
\usepackage{sectsty}
\usepackage{balance}
\usepackage{epstopdf}
\usepackage{graphicx} 
\usepackage{lastpage}
\usepackage[format=plain,justification=raggedright,singlelinecheck=false,font=small,labelfont=bf,labelsep=space]{caption}
\usepackage{fancyhdr}
\begin{document}

\thispagestyle{plain}
\fancypagestyle{plain}{
\renewcommand{\headrulewidth}{1pt}}
\renewcommand{\thefootnote}{\fnsymbol{footnote}}
\renewcommand\footnoterule{\vspace*{1pt}%
\hrule width 3.4in height 0.4pt \vspace*{5pt}}
\setcounter{secnumdepth}{5}

\makeatletter
\def\subsubsection{\@startsection{subsubsection}{3}{10pt}{-1.25ex plus -1ex minus -.1ex}{0ex plus 0ex}{\normalsize\bf}}
\def\paragraph{\@startsection{paragraph}{4}{10pt}{-1.25ex plus -1ex minus -.1ex}{0ex plus 0ex}{\normalsize\textit}}
\renewcommand\@biblabel[1]{#1}
\renewcommand\@makefntext[1]%
{\noindent\makebox[0pt][r]{\@thefnmark\,}#1}
\makeatother
\renewcommand{\figurename}{\small{Fig.}~}
\sectionfont{\large}
\subsectionfont{\normalsize}

\fancyfoot{}
\fancyfoot[R]{\footnotesize{\sffamily{\thepage}}}
\fancyhead{}
\renewcommand{\headrulewidth}{1pt}
\renewcommand{\footrulewidth}{1pt}
\setlength{\arrayrulewidth}{1pt}
\setlength{\columnsep}{6.5mm}
\setlength\bibsep{1pt}

\twocolumn[
  \begin{@twocolumnfalse}
\noindent\LARGE{\textbf{Cold and Slow Molecular Beam}}
\vspace{0.6cm}

\noindent\large{\textbf{Hsin-I Lu,$^{\ast}$\textit{$^{a,b}$} Julia Rasmussen, \textit{$^{c,b}$} Matthew J. Wright, \textit{$^{c,b}$} Dave Patterson, \textit{$^{c,b}$} and
John M. Doyle \textit{$^{c,b}$}}}\vspace{0.5cm}

\noindent \normalsize{Employing a two-stage cryogenic buffer gas cell, we produce a cold, hydrodynamically extracted beam of calcium monohydride molecules with a near effusive velocity distribution. Beam dynamics, thermalization and slowing are studied using laser spectroscopy. The key to this hybrid, effusive-like beam source is a $``$slowing cell$"$ placed immediately after a hydrodynamic, cryogenic source [Patterson \emph{et al}., J. Chem. Phys., 2007, \textbf{126}, 154307]. The resulting CaH beams are created in two regimes. One modestly boosted beam has a forward velocity of $v_f=65$ m/s, a narrow velocity spread, and a flux of $10^9$ molecules per pulse. The other has the slowest forward velocity of $v_f=40$ m/s, a longitudinal temperature of 3.6 K, and a flux of $5\times 10^8$ molecules per pulse.}
\vspace{0.5cm}
 \end{@twocolumnfalse}
  ]

\section{Introduction}

\footnotetext{\textit{$^{a}$~School of Engineering and Applied Sciences, Harvard University, Cambridge, Massachusetts 02138, USA. E-mail: lu@cua.harvard.edu }}
\footnotetext{\textit{$^{b}$~Harvard-MIT Center for Ultracold Atoms, Cambridge, Massachusetts 02138, USA }}
\footnotetext{\textit{$^{c}$~Department of Physics, Harvard University, Cambridge, Massachusetts 02138, USA}}

Physics with cold molecules has recently drawn intense study. Cold and ultracold collisions reveal resonances in the few partial wave regime,\cite{CampbellPRL09,TscherbulJCP08} and quantum threshold laws at cold and ultracold temperatures,\cite{GilijamseSCI06,SawyerPRL08,OspelkausSCI10} as well as new atom-molecule interactions.\cite{CampbellPRL09,GilijamseSCI06}
Cold molecules also provide the opportunity for tests of fundamental laws of physics\cite{HudsonPRL06,VuthaJPB10} and access to cold controlled chemistry with external fields.\cite{KremsPCCP08} In addition, long range and anisotropic electric dipolar interactions can lead to new quantum phases\cite{MicheliNPhys06} and may be applied to quantum computing.\cite{DeMillePRL02}
Future achievements in these research areas are highly dependent on improved production of cold molecules for study.

In general, there are two methods to cool molecules: direct and indirect. Indirect methods have been shown to generate ultracold bi-alkali molecules with high phase space density via assembly of laser cooled atoms. \cite{DanzlNATPHYS10,OspelkausSCI10} On the other hand, direct methods provide access to a diverse set of molecules by directly cooling room-temperature sources, but so far at lower phase space density than the indirect methods.\cite{CarrNJP09} Among the various direct methods, molecular beam techniques play a critical role. Supersonic beams use carrier gases to transfer energetic molecules out of a production area and create translationally cold molecules in the moving frame. However, the high forward velocity of standard supersonic beam methods\cite{PattersonNJP09} limit their utility for some experiments on cold molecules. Additional slowing methods, such as Stark deceleration, Zeeman, and optical slowing,\cite{MeerakkerPRL05,NareviciusPRL08,FultonNPhys06} can further decrease translational energies of molecules in the lab frame. Slow molecules in molecular beams may be extracted by electric, magnetic, or mechanical filtering techniques.\cite{RangwalaPRA03,PattersonJCP07,ArndtEPJD08} Molecules produced purely from filtering techniques, however, are not necessarily internally cold.

Buffer-gas cooling is another direct cooling method.\cite{WeinsteinNAT98,EgorovEPJD04} Buffer-gas cooled beams are applicable to nearly any small molecule\cite{PattersonJCP07,PattersonNJP09,MaxwellPRL05,DeMillePRL09} because only elastic collisions with cold buffer gases are required to translationally and rotationally cool molecules. \cite{MaxwellPRL05,BuurenPRL09} When a buffer-gas beam is operated in the $``$hydrodynamic$"$ enhancement regime, where the diffusion time of molecules is longer than the characteristic time the buffer gas spends in the production cell, molecules can be efficiently extracted into a beam, resulting in high molecular flux.\cite{PattersonJCP07,PattersonNJP09}
Due to collisions with fast, forward-moving buffer gas in a hydrodynamic beam, molecules are accelerated, or $``$boosted$"$, to the forward velocity of the buffer gas, $v_f= \sqrt{2K_B T_{bg}/m_{bg}}$, where $T_{bg}$ and $m_{bg}$ are the temperature and mass of the buffer gas, respectively.\cite{PattersonNJP09,MaxwellPRL05}
Although this results in high molecular fluxes and is useful for many applications, the boosted molecular velocity makes direct loading of high densities of molecules into electromagnetic traps infeasible.
In contrast, buffer-gas beams operated in the diffusive limit have a much lower forward velocity of $v_{f,eff}= \sqrt{2K_B T_{bg}/m_{molecule}}$ and a lower molecular flux.\cite{PattersonJCP07}


Our group previously reported a guided molecular oxygen beam and a slow atomic Ytterbium beam produced from a $``$two-stage cell$"$, which benefits from hydrodynamic extraction, while approaching a near effusive distribution.\cite{PattersonJCP07}
However, the beam dynamics that occurred in the second stage of the cell, which we call the $``$slowing cell$"$, were not fully understood or optimized.
In this letter, we realize cold and slow calcium monohydride (CaH) beams based on a revised two-stage cell design.
Various two-stage cell configurations are investigated, and laser spectroscopy is used to explore the thermalization, slowing, and boosting processes of CaH in the buffer gas beams.
We have achieved a CaH beam with a forward velocity of 65 m/s (longitudinal velocity spread of 40 m/s), $10^9$ molecules per pulse, and a narrow transverse spread of 50 m/s, as well as a beam with a near effusive distribution with a longitudinal temperature of 3.6 K and $\sim$ $5\times10^8$ molecules per pulse.
Such low forward velocities, together with reasonable fluxes, make these CaH beams suitable for both trap loading and cold collision experiments.

\section{Apparatus}

\begin{figure}[h]
\centering
  \includegraphics[height=6 cm]{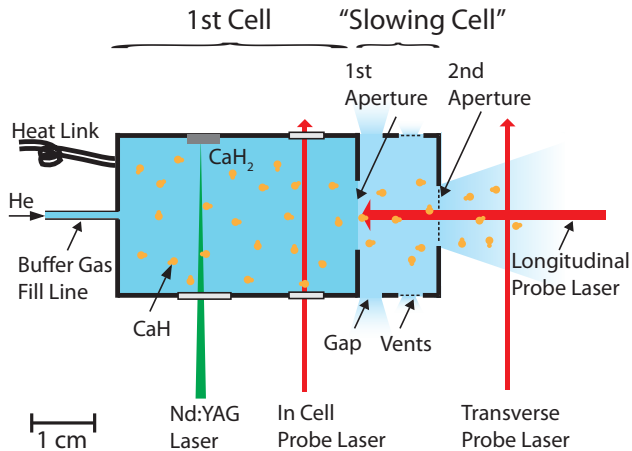}
  \caption{Schematic diagram of a two-stage cell which consists of a single-stage buffer gas cell and a slowing cell. The two cells are separated by a gap and are thermally connected to a helium pump reservoir (not shown) via flexible copper heat links. The buffer gas density in the first cell is maintained by continuously flowing He gas through the fill line. Both the gap between the two cells and two venting ports on the side of the slowing cell reduce the He density in the slowing cell. The venting ports with dimensions 12 mm x 3 mm are covered with 36 \% transparent mesh. A copper mesh covers the final exit aperture, allowing some of the molecules and the buffer gas to pass while only partially reflecting the buffer gas (the molecules can stick to the wires of the mesh). Absorption spectroscopy measurements are performed in the cell and in the transverse beam direction using photodiodes. Fluorescence induced by the longitudinal probe laser is collected by a CCD camera.}
  \label{fgr:combinedapparatus}
\end{figure}

Our experiment is centered around a two-stage copper cell, which consists of a single-stage buffer gas cell and a slowing cell shown in Fig.~\ref{fgr:combinedapparatus}.
The two-stage cell is thermally anchored to a helium pumped reservoir at a base temperature of 1.8 K.
Inside the first cell, CaH ($^2\Sigma^+$) molecules are produced from laser ablation of CaH$_2$ solid precursor with a pulsed Nd:YAG laser, which has a pulse energy of 5 mJ in 4 ns duration. We flow $^4$He continuously into the first cell via a fill line. 
After laser ablation, molecules thermalize with He and are extracted out of the first cell along with continuously flowing buffer gas.

The first cell is 25 mm in inner diameter and 38 mm in length with a 10 mm x 10 mm square aperture in the front.
To achieve a high molecular flux, the dimension of the aperture is chosen to reach hydrodynamic enhancement in molecular beam fluxes. 
For a single-stage cell running in the hydrodynamic limit, the buffer gas density outside its aperture is necessarily high enough for molecules to encounter multiple collisions with the forward-moving buffer gas, leading to increased molecular forward velocities.
To suppress the boosting, the slowing cell, with lower He density, is attached to the first cell.

The slowing cell has internal dimensions of 25 mm diameter and 10 mm length with an exit aperture of 9 mm diameter. The buffer gas density in the slowing cell should be high enough to provide a few collisions to thermalize CaH emitted from the first cell. It should also be low enough to minimize collisions after the exit of the slowing cell, which could introduce an undesirable velocity boost. Two venting ports on the side of the slowing cell are fashioned to leak out some He so as to maintain a proper He density. To further reduce the He density, two cells are separated by a gap, which can be varied between 2.7 - 4.8 mm in distance.

A copper mesh is used to cover the second aperture (on the slowing cell) in an attempt to create a near effusive environment such that molecules inside the slowing cell can experience collisions against the flow direction with He bouncing from the mesh.

The two-stage cell is surrounded by 4-K radiation shields, which are cooled by a pulsed tube refrigerator.
Typical buffer gas flows range from 1-10 standard cubic centimeters per minute (SCCM), corresponding to a He density of $3\times10^{15}-3\times 10^{16}$ cm$^{-3}$ in the first cell.
Activated charcoal on the 4-K shields serves as cryopump to maintain vacuum in the beam area.

CaH in its ro-vibrational ground state is detected via the $A^2 \Pi_{1/2}(v'=0,j'=1/2) \leftarrow X^2\Sigma^+ (v''=0,j''=1/2)$ transition at 695 nm.\cite{DiRosa04} Laser absorption spectroscopy is performed in the first cell to monitor the production yield and thermalization after ablation; it is also used to determine the transverse beam spread and extraction efficiency of CaH into the beam with a transverse probe laser. Forward velocities of the molecular beam are determined by comparing the Doppler shift of fluorescence spectra induced from a longitudinal probe laser relative to the spectrum in the cell.
\section{Results}

\begin{figure}[h]
\centering
  \includegraphics[height=4.4cm]{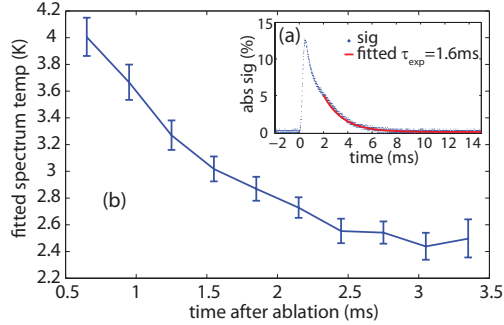}
  \caption{Thermalization dynamics of CaH in the first cell after laser ablation. (a) On resonance absorption signal of CaH. Nd:YAG laser pulse is fired at t=0. Peak absorption level corresponds to $10^{11}$ CaH molecules in the 1st cell. A single exponential decay fit is used to obtain the decay time of CaH when molecules thermalize to the buffer gas.
  (b) Fitted CaH translational temperature as a function of time after laser ablation.}
  \label{fgr:incelldecayfitandtemp}
\end{figure}

Sufficient buffer gas density in the first cell is necessary to ensure thermalization of CaH after laser ablation. The number of collisions $N$ needed for translational temperature $T$ of molecules to become close to $T_{bg}$ is given by $N=-\kappa \ln [(T-T_{bg})/T_i]$, where $T_i$ is the initial temperature of the molecule and $\kappa=(m_{CaH}+m_{He})^2/2m_{CaH}m_{He}$.\cite{DeCarvalhoNJP99}
Thermalizing CaH to within $2T_{bg}$ before molecules hit the cell wall requires a density on the order of $10^{15}$ cm$^{-3}$.
Fig.~\ref{fgr:incelldecayfitandtemp}(a) shows the temporal decay of the CaH absorption signal in the first cell at 2 SCCM flow.
A translational temperature of CaH is obtained by fitting the absorption spectrum with a Gaussian lineshape.
Fitted translational temperatures of CaH for different times after the ablation pulse shown in Fig.~\ref{fgr:incelldecayfitandtemp}(b) indicate that CaH molecules undergo rapid thermalization with the buffer gas and reach a temperature of $\sim$ 2.5 K after 2 ms.
We note that molecules at early times have higher translational energies, possibly contributing to a faster moving molecular beam.

Calibration of the buffer gas density in the first cell helps determine the thermalization process and whether hydrodynamic extraction has been reached. 
Because the ablation laser can desorb superfluid $^4$He from the cell wall, causing a momentary increase in the He density, the measured decay time of molecules $\tau_{exp}$ (see Fig.~\ref{fgr:incelldecayfitandtemp}(a)) is used to directly calibrate the buffer gas density.
The number of CaH molecules in the first cell decays due to diffusion to the cell wall and extraction out of the first cell.
The measured decay time $\tau_{exp}$ is related to the diffusion time $\tau_{diff}$ and the emptying time of the first cell $\tau_{pumpout}$ by a simple relation $\tau_{exp} ^{-1}= \tau_{diff}^{-1}+\tau_{pumpout}^{-1}$, where $\tau_{pumpout}$ can be calculated based on the geometry of the first cell. $\tau_{diff}$ is determined to be 2 ms at 2 SCCM flow, corresponding to a He density of $3\times 10^{15}$ cm$^{-3}$, if the estimated diffusion cross section of $10^{-14}$ cm$^2$ between CaH and He is used. ~\cite{Weinsteinthesis}
This calibrated density is consistent with the He density estimated above to be necessary for thermalization after laser ablation.
For typical buffer gas flows in our experiment (1-10 SCCM), the ratio of $\tau_{diff}$ to $\tau_{pumpout}$ is $0.1-1$, indicating the geometry of the first cell leads to hydrodynamic enhancement in molecular flux.

\begin{figure}[h]
\centering
  \includegraphics[height=4.2cm]{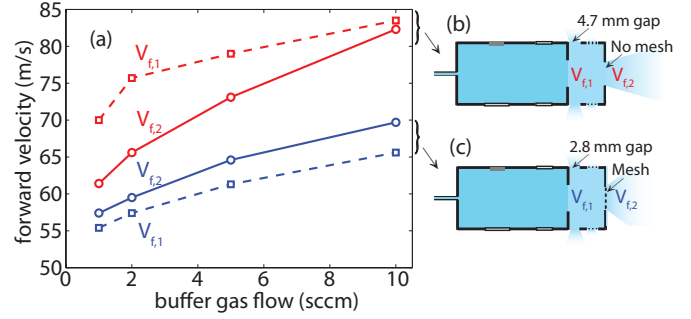}
  \caption{(a) Forward velocities of the molecular beam measured in the gap, $v_{f,1}$, and 3 mm after the second aperture, $v_{f,2}$, as a function of buffer gas flow. Forward velocities plotted in red are for a relatively open cell shown in (b) with 4.7 mm gap and no mesh on the second aperture. (c) shows a different cell configuration with 2.8 mm gap and with mesh covering on the second aperture. Its corresponding forward velocities are plotted in blue in (a).}
  \label{fgr:vfvsbg}
\end{figure}

Once molecules exit the first cell, various factors may determine the final molecular beam velocity: He density mismatch between the two cells, collisions in the slowing cell, collisions with He reflecting from the mesh on the second aperture, and flow dynamics after the second aperture.
We investigate how these factors may affect the molecular beam properties by fine tuning the He density in the slowing cell via varying the gap distance.
Effects of mesh on the second aperture are studied via changing between two different kinds of mesh, as well as uncovering the second aperture completely.
Laser induced fluorescence (LIF) excited by the longitudinal probe laser is collected by a CCD camera, providing information about how beam properties vary spatially. Forward velocity, $v_f$, of the molecular beam is extracted by fitting the LIF spectra with a Gaussian lineshape of the form $f_l (v)\propto e^{-(v-v_f)^2/(0.36 \delta v_l ^2)}$, where $\delta v_l $ is the full width at half maximum (FWHM) of the longitudinal velocity distribution.

Fig.~\ref{fgr:vfvsbg}(a) shows $v_f$ of the molecular beam as a function of buffer gas flow for two different cell configurations as shown in Fig.~\ref{fgr:vfvsbg}(b) and (c).
For one relatively open cell shown in Fig.~\ref{fgr:vfvsbg}(b), forward velocities measured in the gap, $v_{f,1}$, and 3 mm away from the second aperture, $v_{f,2}$, are plotted in red squares and circles, respectively.
This open cell has a 4.7 mm gap with no mesh attached to the second aperture, resulting in the overall open area of the slowing cell to be 16 times larger than the first cell.
For our various cell configurations, the He density in the slowing cell ranges between $10^{14}-10^{15}$ cm$^{-3}$, corresponding to the mean free path of $1-0.1$ cm ($1-10$ collisions) for CaH.
In general, $v_{f,2}$ is lower than $v_{f,1}$, plotted in red, implying molecules can slow down while passing through the slowing cell because of the collisions with the buffer gas. At high flows, $v_{f,2}$ approaches $v_{f,1}$, possibly due to boosting after the second aperture of this open cell.

The other cell configuration shown in Fig.~\ref{fgr:vfvsbg}(c) has a 2.8 mm gap and its second aperture is covered by a coarse copper mesh with a grid size of 750 $\mu$m, 42 \% open area, and 250 $\mu$m thickness. In this case, the total open area of the slowing cell is a factor of 10 larger than the first cell. $v_{f,1}$ for this configuration, plotted in blue squares in Fig.~\ref{fgr:vfvsbg}(a), is roughly 20 m/s lower than that of the previous cell configuration. This indicates boosting of CaH in the gap may be reduced by decreasing the density mismatch between the two cells.
$v_{f,2}$ after the mesh, shown in blue circles in Fig.~\ref{fgr:vfvsbg}(a), only increases slightly and reaches 65 m/s at high flows.
For the cell with mesh, the molecular beam has a typical longitudinal velocity spread of $\delta v_l \sim 40$ m/s and an extraction efficiency of $1-3\%$ based on the transverse beam absorption.
This corresponds to $\sim 10^9$ CaH molecules extracted into the beam per ablation pulse or a peak instantaneous output rate of $\sim 2\times 10^{11}$ molecules per second (given the temporal width of $\sim 5$ ms for the molecular beam pulse).
To extract the FWHM of the transverse velocity distribution, $\delta v_t$, we use a Gaussian lineshape of the form $f_t (v)\propto e^{-v^2/(0.36 \delta v_t ^2)}$, yielding a typical $\delta v_t\sim 50$ m/s.

\begin{figure}[h]
\centering
  \includegraphics[height=4.2cm]{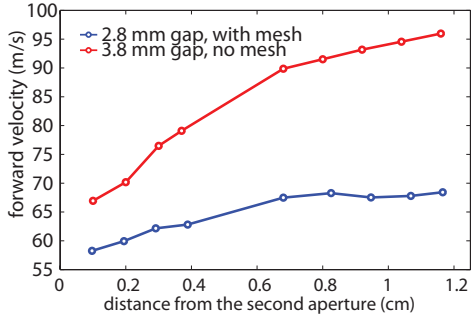}
  \caption{Forward velocity as a function of distance from the second aperture. $v_f$ shown in blue (red) circles is for the cell configuration with 2.8 mm (3.8 mm) gap and with (without) mesh on. Buffer gas flow is 2 SCCM for both data sets. }
  \label{fgr:vfvsdistance}
\end{figure}

To understand how the mesh affects the beam properties, we first analyze the beam dynamics after the second aperture.
In Fig.~\ref{fgr:vfvsdistance}, $v_f$ as a function of distance from the second aperture is plotted in blue circles for the cell shown in Fig.~\ref{fgr:vfvsbg}(b).
$v_f$ plotted in red circles is generated from another open cell with 3.8 mm gap and no mesh on the second aperture, leading to the overall open area of its slowing cell to be 13 times larger than the first cell.
Boosting of the beam should be determined by the number of He-CaH collisions after the second aperture, which is highly dependent on the He density in the beam.
Given the smaller gap, the cell with mesh should have a higher He density in its slowing cell (on the order of $10^{14}$ cm$^{-3}$) than the cell without mesh.
However, $v_f$ for the cell with mesh, as shown in blue circles, increases only modestly by 10 m/s with the increased distance from the second aperture.
In contrast, a beam emitted from an aperture without mesh is rapidly boosted up, as indicated in red circles.
In the presence of mesh, data shown in Fig.~\ref{fgr:vfvsdistance} indicates that molecules experience less boosting and that He-CaH collisions have frozen out within one (2nd) aperture distance.
There is a larger He density drop across the aperture with mesh than that without mesh due to reduced conductance.
In addition, because the mesh grid size is much smaller than the mean free path of He in the slowing cell, each pore on the mesh emits He similar to a point source, resulting in a more divergent He beam profile.
Therefore, we expect the He density in the beam from the aperture with mesh to drop more rapidly than that without mesh.
These two factors result in fewer He-CaH collisions, and hence less boosting, when the aperture is covered with mesh.

\begin{figure}[h]
\centering
  \includegraphics[height=4.6cm]{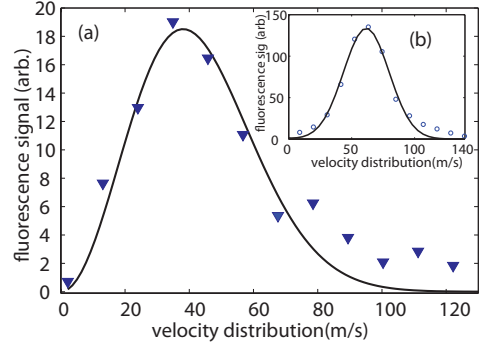}
  \caption{Forward velocity distributions of CaH emitting from (a) a fine mesh with a pore size of 160 $\mu$m and (b) the coarse mesh with a grid size of 750 $\mu$m (b). Both distributions are measured at 3 mm away from the 2nd aperture and at 2 SCCM flow. The solid line in (a) is a fit to an effusive distribution $f_{eff} (v)\propto (v/v_p)^2 e^{-(v/v_p)^2}$, where $v_p=\sqrt{2k_B T/m_{CaH}}$ is the most probable velocity, yielding T=3.6$\pm$0.6 K. The solid line in (b) is a fit to the Gaussian profile $f_l(v)$. }
  \label{fgr:slowbeam}
\end{figure}

We also study how mesh geometry affects the beam dynamics. A fine mesh with 160 $\mu$m pore size, 32 \% open area, and 130 $\mu$m thickness is used to replace the previously described coarse mesh while the gap remains 2.8 mm.
The extraction efficiency of CaH from this fine mesh is a factor of two lower than that of the coarse mesh, corresponding to $\sim 5\times10^8$ molecules per pulse.
However, the forward velocity of this beam is slower, $v_f=$ 40 m/s, as shown in Fig.~\ref{fgr:slowbeam}(a).
The velocity distribution is fitted to an effusive distribution, giving a translational temperature of 3.6 K.
We note that there is a significant molecular population at velocities below 30 m/s (translational energies below 2 K).
As a comparison, Fig.~\ref{fgr:slowbeam}(b) shows the velocity distribution from the coarse mesh. The effusive distribution poorly fits this distribution, possibly because this beam is slightly boosted. The Gaussian profile $f_l(v)$ is used to fit the data, as shown in Fig.~\ref{fgr:slowbeam}(b), giving a forward velocity of $v_f=62$ m/s (translational energy of 9.5 K) and a velocity spread of $\delta v_l=43$ m/s.

Since the pore diameter is comparable to the thickness of this fine mesh, the buffer gas would encounter collisions with the pore wall while passing through the mesh.
The additional slowing from the fine mesh may be understood by the dynamics of He flowing through $``$channels$"$.
If we consider a channel with diameter $D$ and length $L$, its transmission probability $W$ is given by $W= I/I_0 A$, where $I_0$, $I$, and $A$ are the input flux, output flux, and the cross-sectional area of the channel, respectively.\cite{Pauly} Analytic expression of the transmission has been derived \cite{Pauly} and only depends on the ratio of the diameter to the length of the channel, $\beta= D/L$.
In our experiment, the coarse mesh with 750 $\mu$m grid size ($\beta=3$) and the fine mesh with 130 $\mu$m pore size ($\beta=0.8$) should have a transmission of 0.75 and 0.46, respectively.
In other words, more He could be scatted back from the fine mesh than the coarse mesh after reflections from both the pores and physical wires of the meshes are taken into account.
A fine mesh cuts down the extraction efficiency of molecules since CaH would stick to the mesh when colliding with the physical wires.
However, the slow velocity distribution opens up the possibility of direct loading of buffer-gas beams into traps.
\section{Applications}
CaH was the first polar molecule to be trapped inside a magnetic trap via the buffer-gas loading technique. \cite{WeinsteinNAT98}
It has been shown to have good collisional properties with He due to its large rotational splitting and a favorable coupling mechanism of $^2\Sigma$ molecules to structureless atoms,\cite{WeinsteinNAT98,KremsPRA03} making CaH a good candidate for reaching ultracold temperatures via evaporative or sympathetic cooling.
However, pursuing ultracold CaH molecules is impeded in traditional buffer-gas loading experiments because the lingering He atoms desorbing from the cell walls thermally link molecules to their environment.
Although our molecular beam also relies on He buffer gas, CaH can be separated out of the buffer gas through magnetic filtering. \cite{PattersonJCP07} Because of the low translational energies of the beam, our beam may be loaded into magnetic traps without additional slowing.
CaH is also one of the candidates for direct laser cooling of molecules.\cite{DiRosa04,ShumanNAT10}
The higher translational energies molecules possess, the more scattered photons are required for direct laser cooling.
The slow CaH beam reported here is suitable for implementing laser cooling, easing the need for repump lasers.

The slow molecular beam source based on the two-stage cell design is general and can be applied to other molecules, since only elastic collisions with He are involved in the cooling process.
We expect extraction efficiencies to remain on the order of 1 \%.
Radicals with high ablation yields (e.g. CaF, $10^{13}$ per pulse \cite{MaussangPRL05}) naturally increase the output molecular beam flux.
Molecules in gas form at room temperature have been introduced into buffer gas cells via capillary fill lines.\cite{BuurenPRL09,PattersonJCP07}
Avoiding violent discharge processes can potentially lead to molecular beams with slower forward velocities.
\section{Conclusions}
We have demonstrated production of $10^{11}$ CaH molecules per pulse in the ro-vibrational ground state inside the cell. A modestly boosted CaH beam moving at $v_f=$ 65 m/s with a longitudinal velocity spread of $\delta v_l = 40$ m/s and a transverse velocity spread of $\delta v_t = 50$ m/s has $1-3 \%$ extraction efficiency, corresponding to $10^9$ molecules per pulse. A near effusive CaH beam with a translational temperature of 3.6 K has $\sim 0.5 \%$ extraction efficiency, leading to $5\times10^8$ molecules per pulse. This opens the door to direct loading of molecules into magnetic traps.
\section{Acknowledgments}
This work was supported by the Department of Energy and the AFOSR under the MURI award FA9550-09-1-0588.

\footnotesize{
\bibliography{slowbeampaperref} 
\bibliographystyle{rsc} 
}

\end{document}